\documentclass[aps,prc,twocolumn,floatfix,nofootinbib,preprintnumbers,superscriptaddress,longbibliography]{revtex4-1}

\usepackage{epsfig}
\usepackage{graphicx}
\usepackage{stmaryrd}
\usepackage{amssymb,tabularx,dcolumn}
\usepackage{supertabular,ltxtable}
\usepackage{longtable}
\usepackage{amsmath}
\usepackage{amstext}
\usepackage{float}
\usepackage{color}
\usepackage{bm}
\usepackage{CJKutf8}
\usepackage{hyperref}
\usepackage{mathtools}
\usepackage{multirow}
\usepackage{makecell}
\usepackage[utf8]{inputenc}
\usepackage{tabu}
\usepackage{braket}
\usepackage{subfigure}
\usepackage{enumitem}

\makeatletter
\def\@bibdataout@aps{%
\immediate\write\@bibdataout{%
@CONTROL{%
apsrev41Control%
\longbibliography@sw{%
    ,author="08",editor="1",pages="1",title="0",year="1"%
    }{%
    ,author="08",editor="1",pages="1",title="",year="1"%
    }%
  }%
}%
\if@filesw \immediate \write \@auxout {\string \citation {apsrev41Control}}\fi
}
\makeatother

\renewcommand{\vec}{\boldsymbol}

\begin{document}
\begin{CJK*}{UTF8}{gbsn}
\date{today}

\title{Microscopic origin of reflection-asymmetric nuclear shapes}

\author{Mengzhi Chen (陈孟之)}
\affiliation{Department of Physics and Astronomy, Michigan State University, East Lansing, Michigan 48824, USA}
\affiliation{National Superconducting Cyclotron Laboratory, Michigan State University, East Lansing, Michigan 48824, USA}

\author{Tong Li (李通)}
\affiliation{Department of Physics and Astronomy, Michigan State University, East Lansing, Michigan 48824, USA}
\affiliation{National Superconducting Cyclotron Laboratory, Michigan State University, East Lansing, Michigan 48824, USA}

\author{Jacek Dobaczewski}
\affiliation{Department of Physics, University of York, York Y010 5DD, UK}
\affiliation{Institute of Theoretical Physics, Faculty of Physics, University of Warsaw, 02-093 Warsaw, Poland}

\author{Witold Nazarewicz}
\affiliation{Facility for Rare Isotope Beams, Michigan State University, East Lansing, Michigan 48824, USA}
\affiliation{Department of Physics and Astronomy, Michigan State University, East Lansing, Michigan 48824, USA}

\date{\today}

\begin{abstract}
\begin{description}
\item[Background]
The presence of nuclear ground states  with stable reflection-asymmetric  shapes is supported by rich  experimental evidence.  Theoretical surveys of odd-multipolarity  deformations  predict the existence of pear-shaped isotopes  in several fairly localized  regions of the nuclear landscape in the vicinity of  near-lying single-particle shells with $\Delta\ell=\Delta j=3$.

\item[Purpose]
We analyze the role of isoscalar, isovector, neutron-proton, neutron-neutron, and proton-proton multipole interaction energies in inducing the onset of reflection-asymmetric ground-state deformations. 

\item[Methods]
The calculations are performed in the framework of axial reflection-asymmetric Hartree-Fock-Bogoliubov theory  using two Skyrme  energy density functionals and density-dependent pairing force.

\item[Results]
We show that reflection-asymmetric ground-state shapes of atomic nuclei are driven by the  odd-multipolarity neutron-proton  (or isoscalar) part of the nuclear interaction energy. This result is consistent with  the particle-vibration picture, in which  the main driver of octupole instability is the isoscalar octupole-octupole interaction  giving rise to large $E3$ polarizability.

\item[Conclusions]

The necessary condition for  the appearance of localized regions of pear-shaped nuclei in the nuclear landscape is the presence
of parity doublets involving  $\Delta\ell=\Delta j=3$ proton or neutron single-particle shells. This condition alone is, however, not sufficient to determine whether pear shapes actually appear, and -- if so -- what  the corresponding reflection-asymmetric deformation energies are. 
The predicted  small reflection-asymmetric deformation energies result from dramatic 
cancellations between even- and odd-multipolarity components of the nuclear binding energy.

\end{description}
\end{abstract}

\maketitle
\end{CJK*}

\section{Introduction}

While the vast majority of atomic nuclei have either spherical
or ellipsoidal (prolate or oblate) ground-state (g.s.)  shapes, some isotopes
exhibit pear-like shape deformations that intrinsically break reflection symmetry. 
Experimental evidence for such shapes comes from characteristic properties of nuclear spectra, nuclear moments, and electromagnetic matrix elements \cite{Butler1996,Butler2020}. Pear-shaped even-even nuclei display low-energy  negative-parity excitations that are usually attributed to  octupole collective modes. For that reason, pear-shaped nuclei are often referred to as “octupole-deformed.”

There are two regions  of  g.s.\ reflection-asymmetric shapes that have been experimentally established over the years: the neutron-deficient  actinides around $^{224}$Ra and the neutron-rich lanthanides around $^{146}$Ba. Nuclear theory systematically predicts these nuclei to be pear-shaped (see Ref.~\cite{Cao2020} for a recent survey of theoretical results). Other regions of pear-shaped nuclei predicted by theory, i.e., lanthanide nuclei
around $^{200}$Gd as well as actinide and superheavy
nuclei with $184 < N < 206$  are too neutron rich to be accessible
by experiment \cite{Erler2012a,Agbemava2016,Agbemava2017,Xu2017,Cao2020}. 
In general,  deformation energies associated with  reflection-symmetry  breaking shapes are much smaller than those related to  stable ellipsoidal shapes  \cite{Myers1966,Moller2008}. Consequently, 
for  octupole-deformed nuclei,  beyond mean-field methods are needed for a quantitative description, see, e.g., Refs.~\cite{Egido1991,Robledo2016,Xia2017,Fu2018}. 

According to the single-particle (s.p.) picture, the appearance of pear-shaped deformations can be attributed to the  mixing of opposite-parity s.p.\ shells \cite{Strutinsky1956,Lee1957}. In the macroscopic-microscopic (MM) approach, the macroscopic energy favors spherical shapes. Therefore, stable refection-asymmetric shape deformations obtained in the MM method \cite{Nazarewicz1984,Moller2008} can be traced back  to the  shape polarization  originating from proton and neutron s.p.\ levels interacting via parity-breaking fields. Since shell corrections are computed separately for protons and neutrons, the results are usually interpreted in terms of deformation-driving proton or neutron shell effects. The proton-neutron interactions are indirectly considered in the macroscopic energy with the assumption of identical proton and neutron shape deformation parameters, which follow those  of the macroscopic term.

In general, in the description based on the mean-field approach, nuclear shape deformations  result from a coupling between collective surface vibrations of the nucleus 
and  valence nucleons.  Such a
particle-vibration coupling \cite{Bohr52} mechanism can be understood in terms of the nuclear Jahn-Teller effect
\cite{Reinhard1984,Nazarewicz1994}. The tendency towards deformation is  particularly strong
if the Fermi level
lies just between close-lying s.p.\ states. In such a case, the system can become unstable with respect to the mode that couples these states. 
Simple estimates of the particle-vibration  coupling (Jahn-Teller vibronic coupling) for the quadrupole mode
(multipolarity $\lambda=2$) \cite{Bohr75,Bes1975} demonstrate that its contribution to the mass quadrupole moment at low energies doubles the quadrupole moment of  valence nucleons. The Hartree-Fock (HF) analysis  \cite{Dobaczewski1998,Werner1994} confirmed this estimate. It showed that the main contribution to the quadrupole deformation energy comes from the attractive isoscalar quadrupole-quadrupole term, 
which can be well approximated by the neutron-proton  quadrupole interaction.

When it comes to reflection-asymmetric deformations, the leading particle-vibration coupling is the one due to the octupole mode (multipolarity $\lambda=3$). This coupling  generates a vibronic Jahn-Teller interaction between close-lying opposite-parity s.p.\ orbits that may result in a static reflection-asymmetric shape. For g.s.\ configurations of atomic nuclei, such pairs of states  can be found 
just above closed shells and
involve a unique-parity intruder shell $(\ell,j)$ and a normal-parity shell $(\ell-3,j-3)$
around particle numbers  $N_{\rm oct}=34$, 56, 88, and, 134 \cite{Butler1996}. Indeed self-consistent calculations systematically predict pear shapes for nuclei having proton and neutron numbers close to $N_{\rm oct}$.

To understand  the origin of reflection-asymmetric g.s.\ deformations, in this study we extend the quadrupole-energy analysis of 
Refs.~\cite{Dobaczewski1998,Werner1994} to odd-multipolarity shapes. To this end we decompose the total Hartree-Fock-Bogoliubov (HFB) energy 
into isoscalar, isovector, neutron-neutron ($nn$), proton-proton ($pp$), and neutron-proton ($np$) contributions of different multipolarities. 

This paper is organized as follows. In Sec.~\ref{sec:estimates} we estimate the octupole polarizability and coupling strengths of the octupole-octupole interaction.
Section~\ref{sec:multipole} 
describes the multipole decomposition of one-body HFB densities and the HFB energy.
The results of our analysis calculations and an analysis of trends are presented in Sec.~\ref{sec:results}. 
 Finally, Sec.~\ref{sec:summary}  contains the conclusions of this work.

\section{Simple estimate of low-energy octupole coupling}\label{sec:estimates}

In this section, we  follow 
Refs.~\cite{Bohr75,Bes1975}, which used a schematic particle-vibration coupling Hamiltonian consisting of a spherical harmonic-oscillator one-body term and a multipole-multipole residual interaction. This model was used in the early paper \cite{Dobaczewski1998} in the context of quadrupole deformations. 
The model Hamiltonian with the octupole-octupole interaction is 
\begin{equation}\label{eq:qqh}
    \hat{H} = \hat{H}_0 + \frac{1}{2} \kappa_0 \hat{Q}_0\hat{Q}_0 + \frac{1}{2} \kappa_1 \hat{Q}_1\hat{Q}_1,
\end{equation}
where $\hat{Q}_0 = \hat{Q}_n+\hat{Q}_p$ and $\hat{Q}_1 = \hat{Q}_n-\hat{Q}_p$ are single-particle octupole isoscalar and isovector operators, respectively, and $\hat{H}_0$ is a spherical one-body harmonic-oscillator Hamiltonian.
For the case of high-frequency octupole oscillations (giant octupole resonances), 
the coupling constants of the isoscalar and isovector octupole-octupole interactions, $\kappa_0$ and $\kappa_1$, respectively,  can be written as:
\begin{equation}\label{eq:QQ}
\kappa_0=-\frac{4\pi}{7}\frac{M\omega_0^2}{A\langle r^4\rangle}, 
~~\kappa_1=\frac{\pi V_{\rm sym}}{A\langle r^6\rangle},     
\end{equation}
where $\omega_0$ is the oscillator frequency, $V_{\rm sym}$ is the repulsive symmetry potential ($\sim 130$\,MeV), and $M$ is the nucleon mass. 
Since the isovector coupling constant $\kappa_1$ is positive, the g.s.
 neutron and proton deformations are expected to be similar, as assumed in the MM approaches.

Within the Hamiltonian (\ref{eq:qqh}), the g.s.\  octupole polarizability of the nucleus is given by \cite{Bohr75}
\begin{equation}
 \chi_{3,\tau}= -\frac{\kappa_\tau}{\kappa_\tau +C_3^{(0)}},
\end{equation}
where $\tau=0$ or 1 and  $C_3^{(0)}$ is the restoring force parameter. There are two types of octupole modes involving s.p.\ transitions with $\Delta{\cal N}=1$ or 3, where ${\cal N}$ is the principal oscillator quantum number. The corresponding restoring-force parameters are:
\begin{align}
C_3^{(0)}(\Delta{\cal N}=1)& =\frac{16\pi}{21}\frac{M\omega_0^2}{A\langle r^4\rangle},\\
C_3^{(0)}(\Delta{\cal N}=3)&=3C_3^{(0)}(\Delta{\cal N}=1).
\end{align}
By using  the estimate in Ref. \cite{Bohr75}
\begin{equation}\label{eq:identity}
    \frac{V_{\rm sym}}{M\omega_0^2}\approx 2.9 \frac{\langle r^4\rangle}{\langle r^2\rangle},
\end{equation}
one obtains:
\begin{equation}\label{eq:chi-0}
\chi_{3,0}(\Delta{\cal N}=1)=3, ~~~\chi_{3,0}(\Delta{\cal N}=3)=1/3.
\end{equation}
The isovector octupole polarizabilities are obtained in a similar way by assuming a
uniform density distribution:
\begin{equation}\label{eq:chi-1}
\chi_{3,1}(\Delta{\cal N}=1)=-0.78, ~~~\chi_{3,1}(\Delta{\cal N}=3)=-0.54.
\end{equation}

While the collective octupole modes couple the  $\Delta{\cal N}=1$ and 3 transitions, the low-frequency mode is primarily associated with  the $\Delta{\cal N}=1$ excitations. At low energies, associated with nuclear ground states,
 the strength coefficients in
Eq.~(\ref{eq:QQ}) should be renormalized by factors $(1+\chi_{3,\tau})$ to account for the coupling to high-energy octupole collective vibrations. We indicate them by
$\tilde{\kappa}_\tau= (1+\chi_{3,\tau})\kappa_\tau.$ Following Ref.~\cite{Dobaczewski1998}, we rearrange the octupole-octupole Hamiltonian 
into $nn$, $pp$, and $np$ parts with the coupling constants 
\begin{equation}
\tilde\kappa_{nn}=\tilde\kappa_{pp}=\tilde\kappa_0+\tilde\kappa_1,~~\tilde\kappa_{np}=\tilde\kappa_0-\tilde\kappa_1.
\end{equation}
By assuming the average values of octupole polarizabilities 
$\chi_{3,0}\approx 2$ and $\chi_{3,1}\approx -0.4$,
the  ratio of the coupling constants becomes:
\begin{equation}
\frac{\tilde\kappa_{nn}}{\tilde\kappa_{np}}=\frac{\tilde\kappa_{pp}}{\tilde\kappa_{np}}\approx 0.27.
\end{equation}
We can thus conclude
that the octupole-octupole $np$ interaction may indeed be viewed as being
responsible for the development of the octupole deformation.

\section{Multipole expansion of densities and HFB energy}\label{sec:multipole}

In self-consistent mean-field approaches~\cite{Ring1980,Bender2003,Schunck2019} with
energy-density functionals (EDFs) based on two-body functional
generators, the total energy of a nucleus is expressed as:
\begin{equation}
E = \mathrm{Tr} (T\rho) + \tfrac{1}{2} \mathrm{Tr} (\Gamma\rho) + \tfrac{1}{2}\mathrm{Tr} (\tilde{\Gamma}\tilde{\rho}).
\label{eq:total-energy1}
\end{equation}
Here $T$ is the kinetic energy operator, $\Gamma$ and $\tilde{\Gamma}$ are mean fields in particle-hole
(p-h) and particle-particle (p-p) channels, respectively, and $\rho$
and $\tilde{\rho}$ are one-body p-h and p-p density matrices, respectively. (Instead
of using the standard pairing tensor~\cite{Ring1980}, here we use the
``tilde" representation of the p-p density matrix~\cite{Dobaczewski1984}.)
 The mean fields $\Gamma$ and $\tilde{\Gamma}$ are
defined as
\begin{align}
\label{eq:mean-fields1}
T+\Gamma &= \frac{\delta E}{\delta' \rho}, \\
\label{eq:mean-fields2}
\tilde{\Gamma} &= \frac{\delta E}{\delta' \tilde{\rho}},
\end{align}
where $\delta'$ denotes the variation of the total energy that neglects the
dependence of the functional generators on density, that is, the mean
fields~(\ref{eq:mean-fields1}) and (\ref{eq:mean-fields2}) do not
contain so-called rearrangement terms~\cite{Bender2003}.

\subsection{Multipole decomposition}

As observed in Ref.~\cite{Dobaczewski1998},
the density matrices and mean fields can be split into different multipole components as
\begin{subequations}\label{eq:decompose}
\begin{align}
\label{eq:rho_decompose} \rho &= \rho_{[0]} + \rho_{[1]} + \rho_{[2]} + \rho_{[3]} + \ldots \\
\label{eq:rhop_decompose}\tilde{\rho} &= \tilde{\rho}_{[0]} + \tilde{\rho}_{[1]} + \tilde{\rho}_{[2]} + \tilde{\rho}_{[3]} + \ldots, \\
\label{eq:Gamma_decompose} \Gamma &= \Gamma_{[0]} + \Gamma_{[1]} + \Gamma_{[2]} + \Gamma_{[3]} + \ldots \\
\label{eq:Gammap_decompose}\tilde{\Gamma} &= \tilde{\Gamma}_{[0]} + \tilde{\Gamma}_{[1]} + \tilde{\Gamma}_{[2]} + \tilde{\Gamma}_{[3]} + \ldots,
\end{align}
\end{subequations}
where $\rho_{[\lambda]}$, $\tilde{\rho}_{[\lambda]}$, $\Gamma_{[\lambda]}$, and
$\tilde{\Gamma}_{[\lambda]}$ are rank-$\lambda$ rotational components of $\rho$,
$\tilde{\rho}$,  $\Gamma$, and $\tilde{\Gamma}$, respectively. Traces
appearing in Eq.~(\ref{eq:total-energy1}) are invariant with respect
to unitary transformations, and, in particular, with respect to spatial
rotations. Therefore, the traces act like multipolarity filters
projecting the total energy on a rotational invariant. In this way,
when the multipole expansions
(\ref{eq:decompose}) are inserted in
the expression for the total energy (\ref{eq:total-energy1}),
only diagonal terms remain:
\begin{equation}
\label{eq:total-energy}
E =  E_{[0]} + E_{[1]} + E_{[2]} + E_{[3]} + \ldots,
\end{equation}
where
\begin{equation}
\label{eq:E_decompose_prl}
E_{[\lambda]} = \tfrac{1}{2} \mathrm{Tr} (\Gamma_{[\lambda]}\rho_{[\lambda]}) + \tfrac{1}{2}\mathrm{Tr} (\tilde{\Gamma}_{[\lambda]}\tilde{\rho}_{[\lambda]}).
\end{equation}
In the above equation, we add the kinetic energy to the monopole energy
$E_{[0]}$ since $T$ is a scalar operator which implies $E_{\rm kin}=\mathrm{Tr} (T\rho)\equiv\mathrm{Tr} (T\rho_{[0]})$.
Therefore we define
\begin{equation}
E_{[0]} = E_{\rm kin} + \tfrac{1}{2} \mathrm{Tr} (\Gamma_{[0]}\rho_{[0]}) + \tfrac{1}{2}\mathrm{Tr} (\tilde{\Gamma}_{[0]}\tilde{\rho}_{[0]}).
\end{equation}

When parity symmetry is conserved, only even-$\lambda$ multipolarities appear in
Eqs.~(\ref{eq:decompose}) and (\ref{eq:total-energy}). In
Refs.~\cite{Dobaczewski1998,Werner1994}, this allowed for
analyzing the monopole ($\lambda=0$), quadrupole ($\lambda=2$), and higher
even-$\lambda$ components. In the present work, we analyze broken-parity
self-consistent states and focus on the reflection-asymmetric  (odd-$\lambda$) components
of the expansion. As our multipole expansion is defined with
respect to the center of mass of the nucleus, the integral of the isoscalar dipole density
$\rho_{[1]}$, namely, the total isoscalar dipole moment, vanishes by construction. Nevertheless, the dipole density
$\rho_{[1]}$ and dipole energy $E_{[1]}$ can still be nonzero.

In the spherical s.p.\ basis,
the expansions~(\ref{eq:decompose}) can
be realized by the angular-momentum coupling of basis wave
functions. Since the HFB equation is usually solved in a
deformed basis, an explicit basis transformation is then
needed. Moreover, the direct angular-momentum coupling does not
benefit from the fact that Skyrme EDFs only depend on
(quasi)local densities, which is the property that greatly simplifies
the HFB problem. Inspired by the latter observation, in this work, we
determine the multipole expansions of (quasi)local densities and
(quasi)local mean fields directly in the coordinate space.

With axial symmetry assumed, particle density $\rho(\vec{r})$ can be decomposed as \cite{Vautherin1973}
\begin{equation}
\label{eq:density-expansion1}
\rho(\vec{r}) = \sum_{J} \rho_{[\lambda]}(r) Y_{J,M=0}(\Omega) ,
\end{equation}
where
\begin{equation}
\label{eq:density-expansion2}
\rho_{[\lambda]}(r) = \int d\Omega \rho(\vec{r}) Y^*_{J,M=0}(\Omega).
\end{equation}
An identical decomposition can be carried out for all isoscalar
($t=0$) and isovector ($t=1$) (quasi)local p-h
densities~\cite{Perlinska2004}
$\varrho_t\equiv\{\rho_t$, $\tau_t$, $
\Delta \rho_t $, $\mathbb{J}_t, \nabla\cdot\mathbf{J}_t\}$, plus local neutron ($q=n$) and proton ($q=p$) pairing densities $\tilde{\rho}_q$. 
The p-h densities depend on neutron and proton densities in the usual way:
\begin{equation}
\label{eq:densities-np}
\varrho_0 =\varrho_n+\varrho_p,~~\varrho_1=\varrho_n-\varrho_p.
\end{equation}
Our
strategy is  to use the energy-density expression for the time-even total
energy (\ref{eq:total-energy1}),
\begin{equation}
\label{eq:total-energy2}
E = \int\mathrm{d}^3\vec{r}\left\{\frac{\hbar^2}{2m}\tau_0(\vec{r})
+\mathcal{H}(\vec{r})+\tilde{\mathcal{H}}(\vec{r})\right\},
\end{equation}
where the standard Skyrme energy densities read~\cite{Perlinska2004,Engel1975}:
\begin{subequations}
\begin{align}
\mathcal{H}(\vec{r}) &= \sum_{t=0,1} \mathcal{H}_t(\vec{r}), \\
\tilde{\mathcal{H}}(\vec{r}) &= \sum_{q=p,n} \tilde{\mathcal{H}}_q(\vec{r}),
\end{align}
\end{subequations}
and where
\begin{subequations}\label{eq:energydensity}
\begin{align}
        \mathcal{H}_t(\vec{r}) &=
        C^{\rho}_t\rho_t^2(\vec{r}) + C^{\Delta\rho}_t
        \rho_t(\vec{r})\Delta\rho_t(\vec{r}) 
        \nonumber \\
     &   \quad + C^{\tau}_t\rho_t(\vec{r})\tau_t(\vec{r})
        +C^{\text{J}}_t\mathbb{J}_t^2(\vec{r}) \label{eq:energy-density-p-h}
     \\
  &    \quad + C^{\nabla J}_t\rho_t(\vec{r}) \nabla\cdot\mathbf{J}_t(\vec{r})   \nonumber  , \\
\label{eq:energy-density-p-p}
        \tilde{\mathcal{H}}_q(\vec{r}) &= \tfrac{1}{4}V_q\left[1-
        V_1\left(\frac{\rho(\vec{r})}{\rho_0}\right)^{\gamma}\right]\tilde{\rho}_q^2(\vec{r}).
\end{align}
\end{subequations}
For simplicity, the Coulomb energy is not included in Eq.~(\ref{eq:total-energy2}); it will be discussed later. 

It is convenient to rewrite
the energy densities (\ref{eq:energydensity})  in terms of local p-h and p-p
potentials as
\begin{subequations}\label{eq:energydensity1}
\begin{align}
        \mathcal{H}_t(\vec{r}) &=
        V_t(\vec{r})\rho_t(\vec{r})
                    + \sum_{ij} \mathbb{V}_{tij}(\vec{r})\mathbb{J}_{tij}(\vec{r}), \label{eq:Htau} \\
        \tilde{\mathcal{H}}_q(\vec{r}) &= \tilde{V}_q(\vec{r})\tilde{\rho}_q(\vec{r}),
\end{align}
\end{subequations}
where
\begin{subequations}\label{eq:potentials}
\begin{align}
        V_t(\vec{r}) &=
        C^{\rho}_t\rho_t(\vec{r}) + C^{\Delta\rho}_t
        \Delta\rho_t(\vec{r}) \nonumber \\
\label{eq:potential-p-h1}
        & \hspace*{0.5cm}+ C^{\tau}_t\tau_t(\vec{r})
        +C^{\nabla J}_t \nabla\cdot\mathbf{J}_t(\vec{r}) , \\
\label{eq:potential-p-h2}
        \mathbb{V}_{tij}(\vec{r}) &= C^{\text{J}}_t\mathbb{J}_{tij}(\vec{r}) , \\
\label{eq:potential-p-p}
        \tilde{V}_q(\vec{r}) &= \tfrac{1}{4}V_q\left[1-
        V_1\left(\frac{\rho(\vec{r})}{\rho_0}\right)^{\gamma}\right]\tilde{\rho}_q(\vec{r}),
\end{align}
\end{subequations}
with indices $i,j$ denoting the components of the spin-current tensor density $\mathbb{J}_{tij}(\vec{r})$
in three dimensions. In analogy to Eqs.~(\ref{eq:density-expansion1})
and (\ref{eq:density-expansion2}), we then determine the multipole
expansions of the local potentials
(\ref{eq:potentials}). In this way, the
total energy (\ref{eq:total-energy}) can be decomposed into multipole
components:
\begin{widetext}
\begin{eqnarray}
\label{eq:E_decompose_prc}
E_{[\lambda]} &=&\!\!  \int\!\!\mathrm{d}^3\vec{r}\!
\left[\sum_{t=0,1}\left\{V_{t[\lambda]}(\vec{r})\rho_{t[\lambda]}(\vec{r})
    + \sum_{ij} \mathbb{V}_{tij[\lambda]}(\vec{r})\mathbb{J}_{tij[\lambda]}(\vec{r})\right\}
    + \sum_{q=p,n}\tilde{V}_{q[\lambda]}(\vec{r})\tilde{\rho}_{q[\lambda]}(\vec{r})\right].
\end{eqnarray}
\end{widetext}

Finally, the same strategy can be applied to the Coulomb energy,
which contributes to the multipole terms of
Eq.~(\ref{eq:total-energy}) through the multipole expansions of
direct and exchange potentials:
\begin{equation}
        E^{\text{Coul}}_{[\lambda]} = \int \mathrm{d}^3 \vec{r}
        \left[\tfrac{1}{2}V^{\text{dir}}_{[\lambda]}(\vec{r}) +
        \tfrac{3}{4}V^{\text{exc}}_{[\lambda]}(\vec{r})\right]\rho_{p[\lambda]}(\vec{r}),
\end{equation}
where
\begin{eqnarray}
        V^{\text{dir}}(\vec{r}) &=& e^2\int \mathrm{d}^3 \vec{r'}
        \frac{\rho_p(\mathbf{r'})}{|\vec{r}-\vec{r'}|}, \\
        V^{\text{exc}}(\vec{r}) &=& -e^2\left[\tfrac{3}{\pi}\rho _p(\vec{r})\right]^{\tfrac{1}{3}}.
\end{eqnarray}

\subsection{Isospin and neutron-proton energy decomposition}

In the isospin scheme, the total energy can be written as
\begin{equation}
E=E^{t=0} +  E^{t=1} 
+E^{\text{Coul}} +
E^{\text{pair}},
\end{equation}
where
\begin{subequations}
\begin{align}\label{eq:Et}
 E^t &=  E_{\rm kin} \,\delta_{t0}+ \int \mathrm{d}^3 \vec{r} \mathcal{H}_t(\vec{r}), \\
 E^{\rm pair} &= \sum_{q=p,n} \int \mathrm{d}^3 \vec{r} \tilde{\mathcal{H}}_q(\vec{r}).
\end{align}
\end{subequations}
 Note that the kinetic energy $E_{\rm kin}$ is included in the isoscalar energy $E^{t=0}$. The Coulomb energy  $E^{\text{Coul}}$ is separated out
because the Coulomb interaction breaks the isospin symmetry. 
The pairing functional is not isospin invariant either 
as the neutron and proton pairing strengths differ.         

By decomposing the isoscalar and isovector  p-h densities  $\varrho_t$ into the neutron and proton components (\ref{eq:densities-np}), 
the total energy can be expressed in the 
neutron-proton scheme~\cite{Dobaczewski1998}:
\begin{equation}\label{Eq:Enp}
E=E_{\rm kin} + E^{nn} + E^{pp} + E^{np}.
\end{equation}
In Eq.~(\ref{Eq:Enp}), the individual $E^{qq'}$ components ($q,q'=n$ or $p$):
\begin{equation}\label{Eq:Hnp}
 E^{qq'} =\int \mathrm{d}^3 \vec{r} \left[\mathcal{H}_{qq'}(\vec{r}) +\delta_{qq'}\tilde{\mathcal{H}}_{q}(\vec{r})\right],
\end{equation}
are defined through  the energy densities $ \mathcal{H}_{qq'}$ and
$\tilde{\mathcal{H}}_{q}$, which  are bilinear in the densities  $\varrho_q$ or $\tilde{\rho}_q$.
Note that 
the Coulomb energy $E^{\text{Coul}}$ is included in the proton energy $E^{pp}$.
As discussed earlier, all  the energy terms entering the isospin and neutron-proton  decompositions
can be  expanded into  multipoles.

\section{Results}\label{sec:results}
The systems we studied are even-even barium, radium and uranium isotopes. 
They are predicted to have stable pear shapes   at certain neutron numbers \cite{Cao2020}.
For comparison, we also calculate ytterbium isotopes which have stable quadrupole but no reflection-asymmetric deformations.
We performed axial HFB calculations  using
the code {\sc hfbtho} (v3.00) \cite{Perez2017}
for two Skyrme EDFs given by 
SLy4~\cite{Chabanat1998} and UNEDF2~\cite{Kortelainen2014} parametrizations. We used the mixed-pairing
strengths of $V_n=-325.25$\,MeV and $V_p=-340.06$\,MeV (SLy4) and $V_n=-231.30$\,MeV and $V_p=-255.04$\,MeV (UNEDF2).
For UNEDF2, we did not apply the Lipkin-Nogami treatment of pairing; instead, we took the neutron pairing strength $V_n$ to reproduce the average experimental neutron pairing gap for $^{120}$Sn, $\Delta_n$ = 1.245 MeV. The proton pairing strength $V_p$ was adjusted proportionally based on the default values of $V_n$ and $V_p$.

In the first step, we performed parity-conserving calculations
by constraining the octupole deformation to zero and determined the
corresponding equilibrium quadrupole deformation $\beta_2^{(0)}$.
At the fixed value of $\beta_2^{(0)}$, we varied
$\beta_3$ from 0.0 to 0.25. In the {\sc hfbtho} code, multipole constraints are actually applied to quadrupole ($Q_{20}$) and octupole ($Q_{30}$) moments related to
 $\beta_2$ and $\beta_3$ through
\begin{equation}
\begin{aligned}
        \beta_2 &=Q_{20}/\left(\sqrt{\frac{16\pi}{5}}
        \frac{3}{4\pi}AR_0^2\right),\\
        \beta_3 &=Q_{30}/\left(\sqrt{\frac{16\pi}{7}}\frac{3}{4\pi}AR_0^3\right),
\end{aligned}
\end{equation}
where $A$ is the mass number, $R_0=1.2$\,fm$\times A^{1/3}$, and 
\begin{equation}
\begin{aligned}
Q_{20}&=\left\langle 2 z^{2}-x^{2}-y^{2}\right\rangle, \\
Q_{30}&=\left\langle z\left(2 z^{2}-3 x^{2}-3 y^{2}\right)\right\rangle.
\end{aligned}
\end{equation}

\begin{figure}[htbp]
\includegraphics[width=1.0\linewidth]{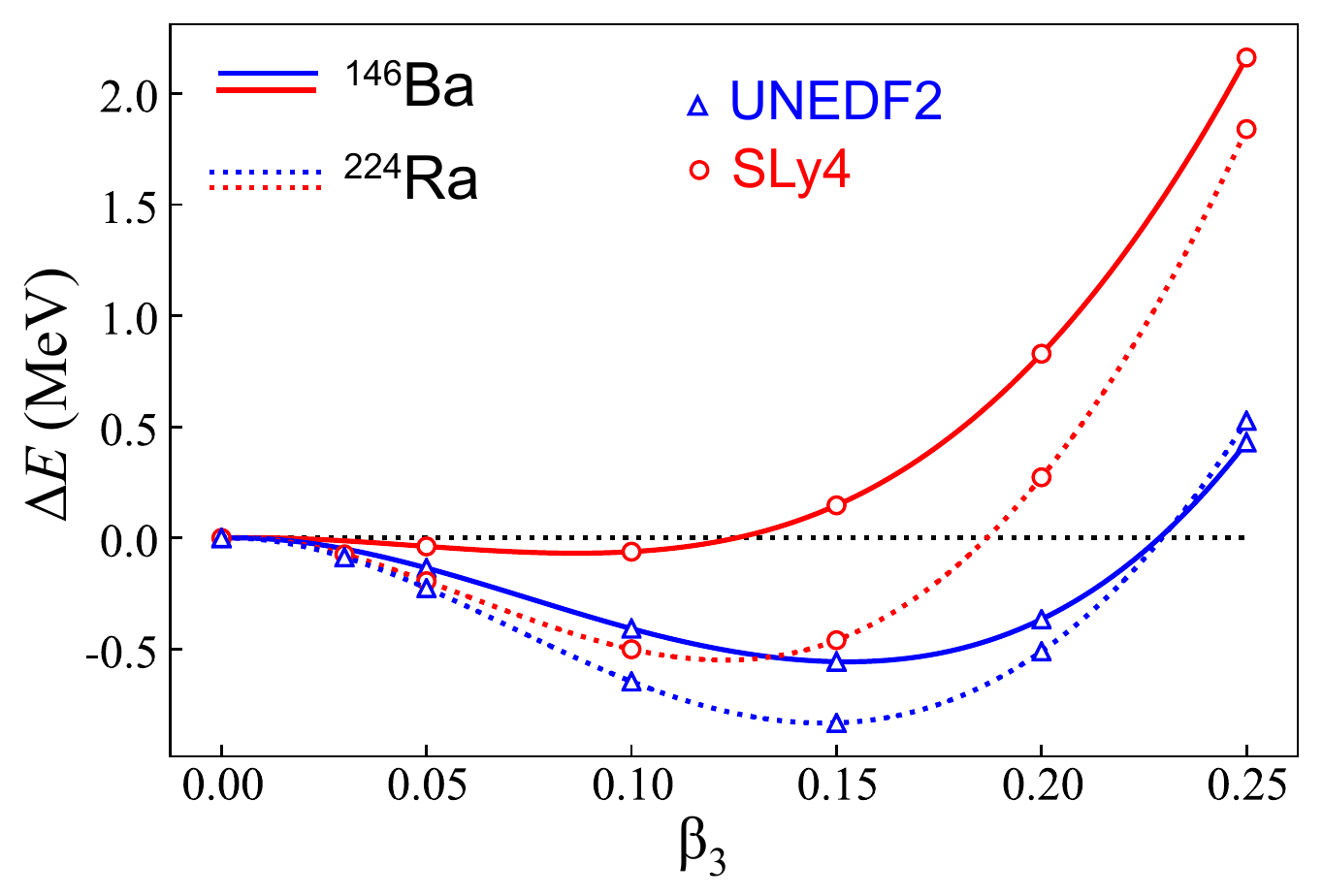}
   \caption{The deformation energies,
        $\Delta E(\beta_3)=E(\beta_3) - E(\beta_3=0)$,
        as functions of $\beta_3$ for $^{224}$Ra (dashed lines) and $^{146}$Ba (solid lines) calculated at
       $\beta_2^{(0)}$ with the  SLy4 (circles) and UNEDF2 (triangles)  EDFs.}
        \label{fig:E_diff}
\end{figure}
 Figure~\ref{fig:E_diff}
shows reflection-asymmetric  deformation energies  $\Delta E(\beta_3)=E(\beta_3) - E(\beta_3=0)$ determined for $^{224}$Ra and  $^{146}$Ba obtained in this way.
We see that  UNEDF2 gives a higher octupole deformability than SLy4
in both nuclei. This is consistent with the results of Ref.~\cite{Cao2020}.

\subsection{Multipole expansion of the deformation energy}

\begin{figure}[htbp]
\includegraphics[width=1.0\linewidth]{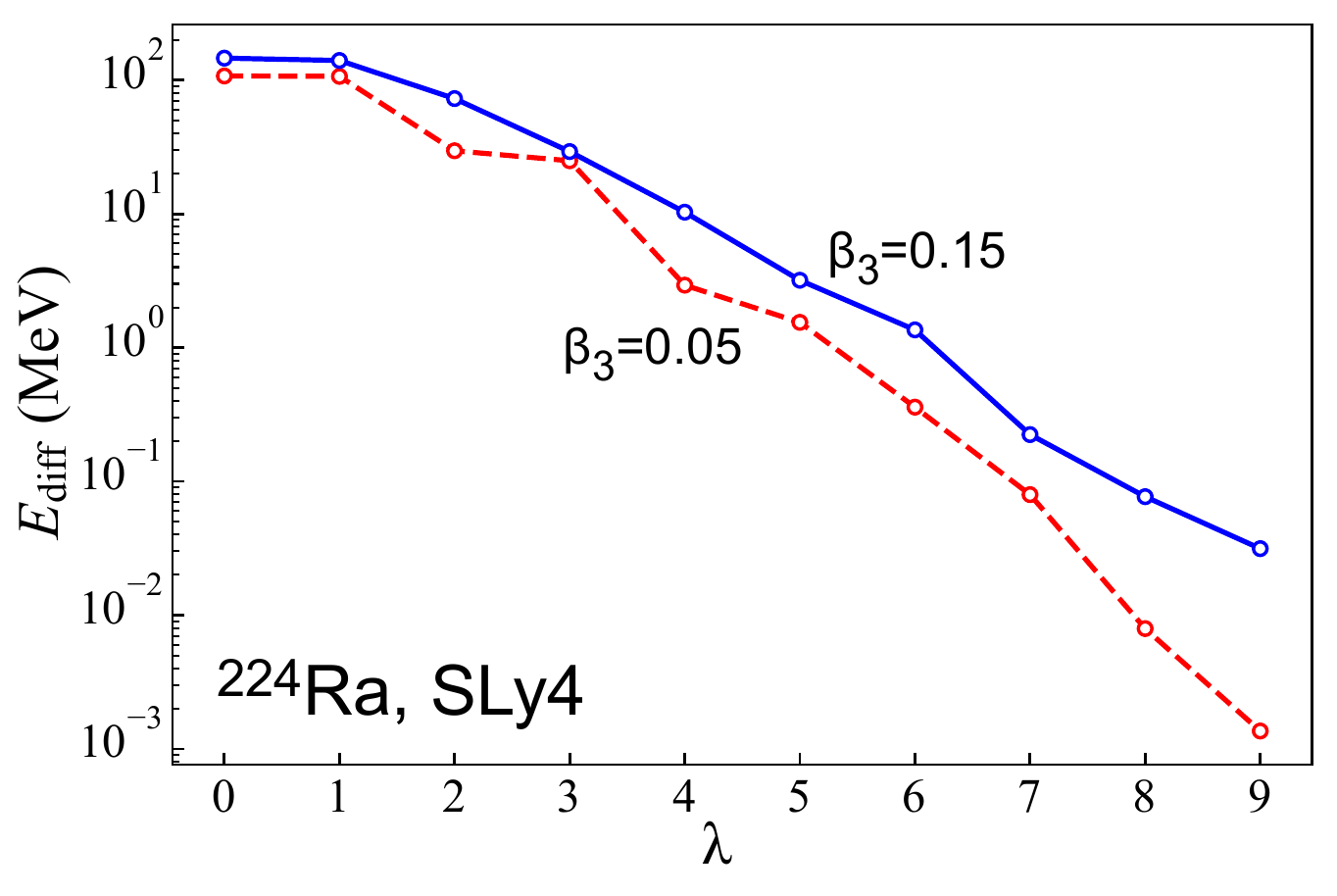}
\caption{Convergence of $E_{\text{diff}}(\lambda)$
 (\ref{eq:expansion1})  for $^{224}$Ra computed with SLy4 at $\beta_3$=0.05 (dashed line) and 0.15 (solid line).}
\label{fig:convergence_Ra224}
\end{figure}
The convergence of the  multipole expansion (\ref{eq:total-energy})  provides a check on   the accuracy of our results. In Fig.~\ref{fig:convergence_Ra224}, we show the energy difference, 
\begin{equation}\label{eq:expansion1}
E_{\text{diff}}(\lambda)=\sum_{\lambda'=0}^{\lambda} E_{[\lambda']}-E
\end{equation}
for  $^{224}$Ra at two values of the
octupole deformation, $\beta_3=0.05$ and 0.15.
We see that at $\beta_3=0.15$, the multipole components
decrease exponentially with $\lambda$, with the monopole component off by about 150\,MeV
and the sum up to $\lambda=9$ exhausted up to about 20\,keV.
At a small octupole deformation of $\beta_3=0.05$, high-order contributions
decrease. As expected, the octupole component brings now less
energy as compared to the quadrupole one. The results displayed in Fig.~\ref{fig:convergence_Ra224} convince us that cutting the multipole expansion of energy at $\lambda=9$ provides sufficient accuracy.

%
\begin{figure}[htb]
\includegraphics[width=1.0\linewidth]{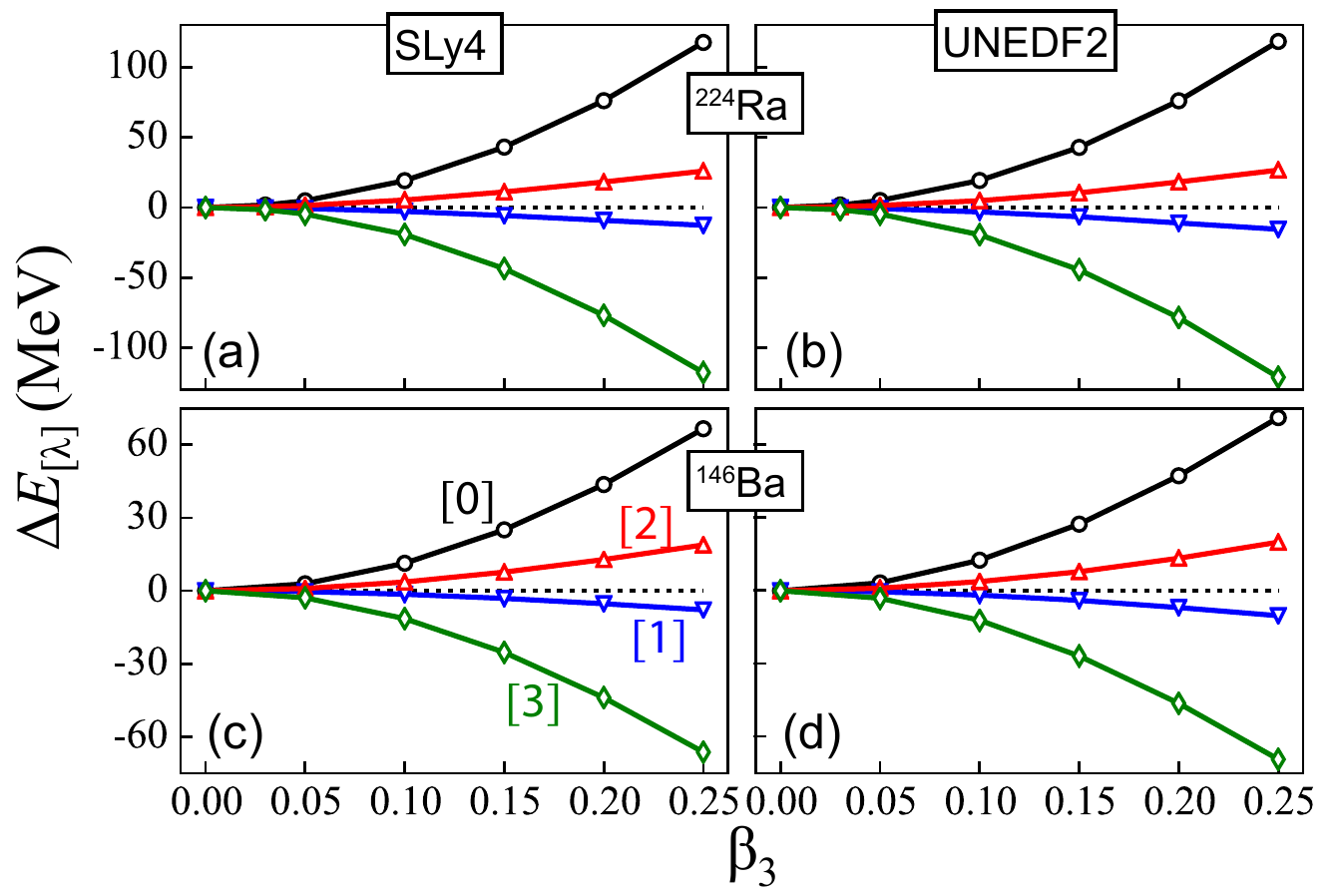}
\caption{Multipole components, $\Delta E_{[\lambda]}(\beta_3)=E_{[\lambda]}(\beta_3) - E_{[\lambda]}(\beta_3=0)$,
of the total deformation
energy shown in Fig.~\protect\ref{fig:E_diff}, plotted for $\lambda=0-3$ 
as functions of the octupole deformation $\beta_3$ at $\beta_2^{(0)}$.
Upper (lower) panels show results for $^{224}$Ra ($^{146}$Ba)
obtained with the  SLy4 (left) and  UNEDF2 (right) EDFs.}
\label{fig:E_diff_order}
\end{figure}
%
Figure~\ref{fig:E_diff_order} shows how the reflection-asymmetric deformation energy builds up. 
It presents the  four leading multipole components $\Delta E_{[\lambda]}(\beta_3)=E_{[\lambda]}(\beta_3) - E_{[\lambda]}(\beta_3=0)$, for $\lambda=0-3$, 
of the  deformation energies shown in Fig.~\ref{fig:E_diff}. We can see
that the pattern of contributions of different multipolarities is fairly generic: 
it weakly depends on the choice of the nucleus or EDF.
Figure~\ref{fig:E_diff_order} clearly demonstrates that the main driver of reflection-asymmetric shapes is  a  strong attractive octupole energy $\Delta E_{[3]}$. 
The attractive dipole energy $\Delta E_{[1]}$ is much weaker.
The  monopole and quadrupole energies are repulsive along the trajectory of $\beta_3$ (with a fixed quadrupole deformation $\beta_2^{(0)}$) and essentially cancel the octupole contribution. Indeed, 
one can  note that while 
individual multipole components can be of the order of tens of MeV,  the
total reflection-asymmetric  deformation energy shown in Fig.~\ref{fig:E_diff} is  an order
of magnitude smaller. Therefore, the final reflection-asymmetric correlation results from a large
cancellation between individual multipole components, and even a relatively small
variation of  any given   component can significantly shift the net result. In addition, as discussed in Sec.~\ref{sec:Ndependence} below, 
higher-order multipole components ($\lambda>3$) can be important for the total energy balance.

\subsection{Isospin and neutron-proton structure of the octupole deformation energy}

\begin{figure}[htb]
\includegraphics[width=1.0\linewidth]{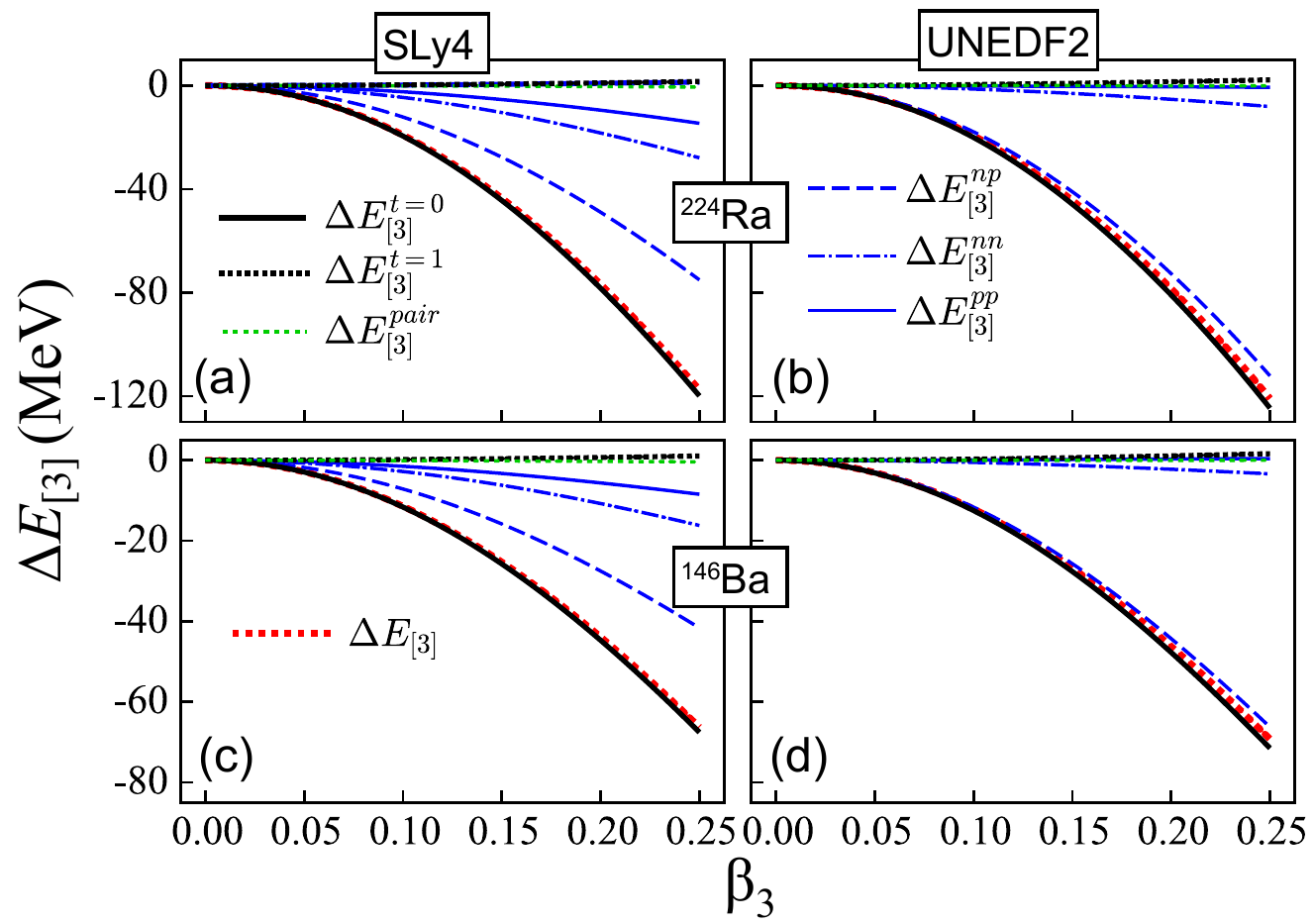}
\caption{Similar to  Fig.~\protect\ref{fig:E_diff_order} but for different isospin and neutron-proton components of the octupole energy $\Delta E_{[3]}$.}
\label{fig:E_diff_iso}
\end{figure}
To analyze the origin of the octupole energy $\Delta E_{[3]}$,
in Fig.~\ref{fig:E_diff_iso} we show its  isospin and neutron-proton components as defined 
 in Eqs.~(\ref{eq:Et}) and (\ref{Eq:Hnp}). Again, a generic pattern  emerges.
In all cases, the octupole energy is almost equal to its isoscalar 
part $\Delta E^{t=0}_{[3]}$. The isovector energy  $\Delta E^{t=1}_{[3]}$
is indeed very small, even if the studied nuclei have a significant
neutron excess; this is consistent with the simple estimates of Sec.~\ref{sec:estimates}. The contribution from the pairing energy  $\Delta E^{\rm pair}_{[3]}$ is also practically negligible. In the neutron-proton scheme, the $np$ component always clearly dominates
the $nn$ and $pp$ terms. The latter two are very small for UNEDF2 and hence
$\Delta E_{[3]} \approx \Delta E^{t=0}_{[3]}\approx \Delta E^{np}_{[3]}$ for this EDF.
For SLy4, the $nn$ and $pp$ terms provide larger contributions to the octupole deformation energy, accompanied by a reduction of the $np$ term. Regardless of these minor differences between the EDFs, 
we can safely conclude that it is the isoscalar octupole 
component (or the  $np$ octupole energy component)  that plays the dominant
role in building up the nuclear octupole deformation.

\subsection{Reflection-asymmetric deformability along Isotopic chains}\label{sec:Ndependence}

At this point,  we are ready to study structural changes  that dictate the appearance of
nuclear reflection-asymmetric deformations.  The results shown in Figs.~\ref{fig:E_diff_order}
and~\ref{fig:E_diff_iso} tell us that a mutual cancellation of
near-parabolic shapes of different components of the deformation
energy results in a clearly non-parabolic dependence of the total deformation energy, as
seen in Fig.~\ref{fig:E_diff}. Therefore, to track back the positions
and energies of the equilibrium reflection-asymmetric deformations to the properties
of specific interaction  components is not easy. To this end, we
analyze the properties of reflection-asymmetric deformabilities of nuclei, that is, we
concentrate on the curvature of reflection-asymmetric deformation energies at
$\beta_3=0$. 
\begin{figure}[htbp]
\includegraphics[width=1.0\linewidth]{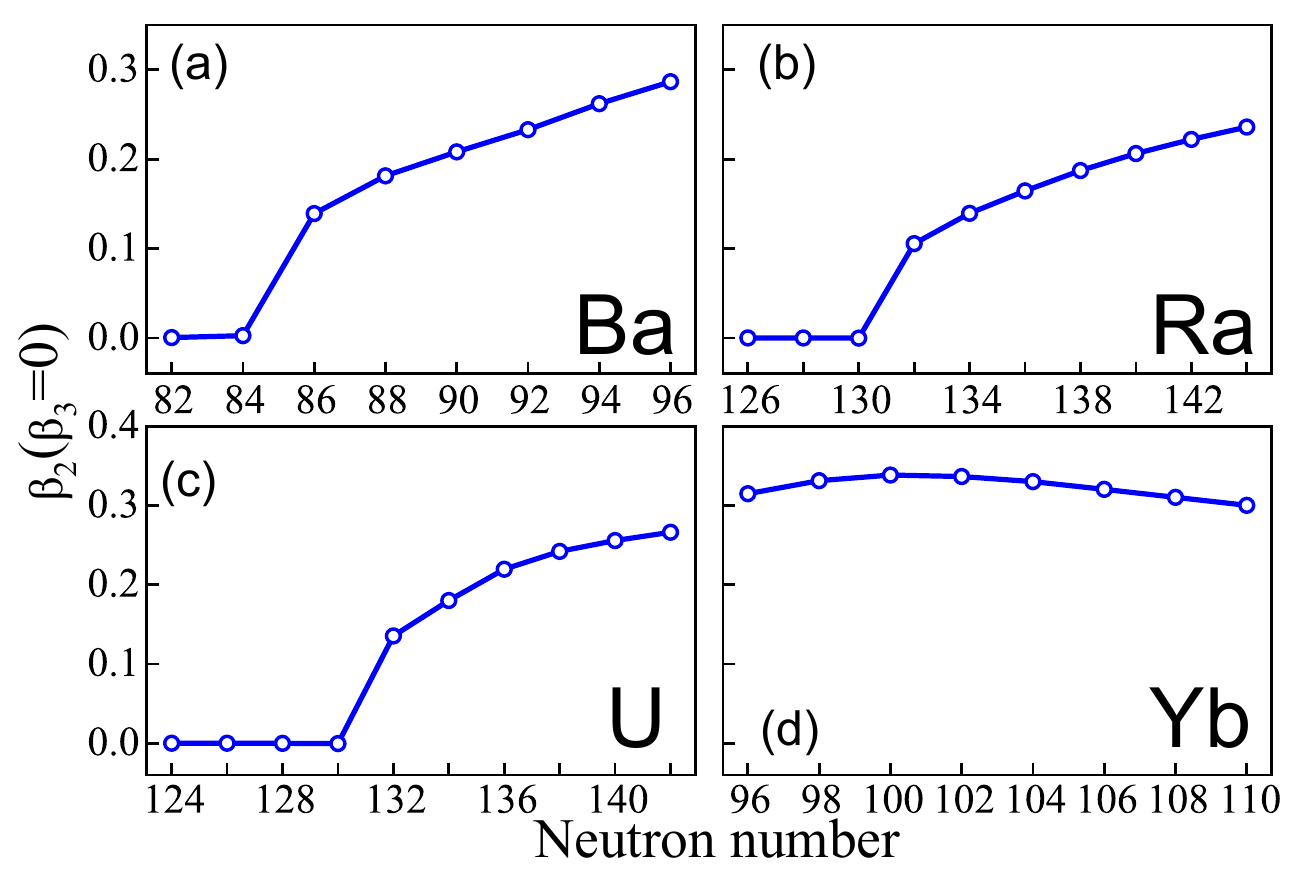}
\caption{Equilibrium quadrupole deformations $\beta_2^{(0)}$  as functions of  $N$ for the isotopic chains of (a) Ba, (b) Ra, (c) U, and (d) Yb computed with the SLy4 EDF.
}
\label{fig:3_isotopes}
\end{figure}
To investigate the variation of the reflection-asymmetric deformability
with neutron number,
we performed SLy4-HFB calculations 
for the isotopic chains of even-even $^{138-152}$Ba, 
$^{214-232}$Ra, and $^{216-234}$U isotopes, which are in the region of  reflection-asymmetric instability, as well as $^{166-180}$Yb, which are expected to be reflection-symmetric \cite{Cao2020}.
In Fig.~\ref{fig:3_isotopes} we show the
baseline quadrupole deformations $\beta_2^{(0)}$.
For the Ba, Ra, and U isotopic chains, spherical-to-deformed shape transitions are predicted slightly above the neutron magic numbers.
The  considered open-shell Yb isotopes are all predicted to be well deformed.

\begin{figure}[htbp]
\includegraphics[width=1.0\linewidth]{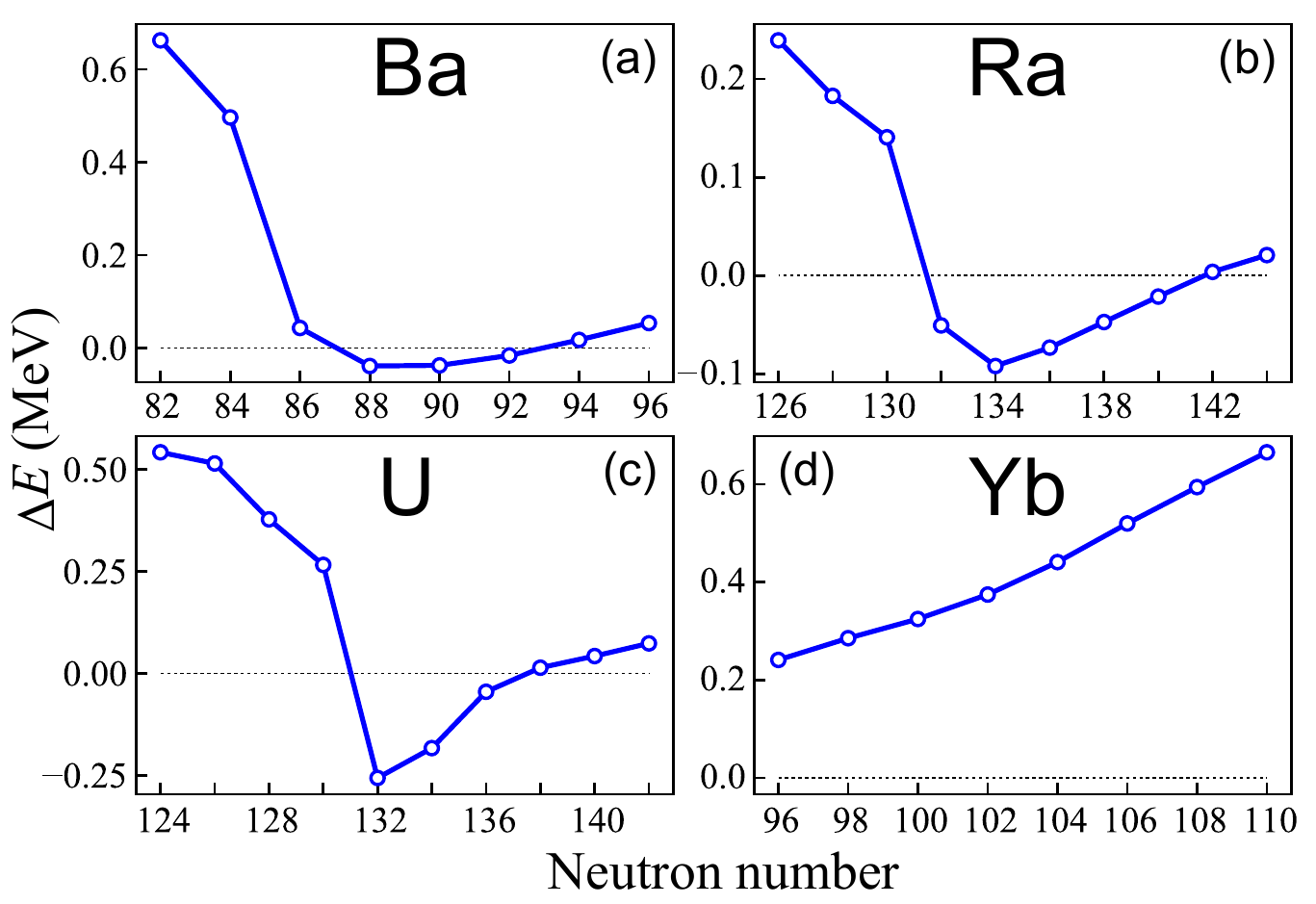}
\caption{Similar to Fig.~\protect\ref{fig:3_isotopes} but for the deformation energy
 $\Delta E = E(\beta_3=0.05) - E(\beta_3=0).$
}
\label{fig:3_isotopes_beta3}
\end{figure}
As a quantitative measure of the octupole deformability, we analyze
the deformation energy $\Delta E = E(\beta_3=0.05) - E(\beta_3=0)$ calculated at a small octupole deformation of $\beta_3=0.05$, with the quadrupole deformation fixed at $\beta_2^{(0)}$. 
We have checked that for different energy components, curvatures $\Delta E/\beta_3^2$ are stable within about 1\% up to $\beta_3=0.05$, so
values of $\Delta E$ taken at $\beta_3=0.05$ constitute valid measures of the
octupole stiffness. In Fig.~\ref{fig:3_isotopes_beta3} we show the values of $\Delta E$
calculated for the four studied isotopic chains. We see that the negative
values of $\Delta E$ delineate regions of neutron numbers where
reflection-asymmetric  deformations set in in Ba, Ra, and U isotopes \cite{Cao2020}.

\begin{figure}[htbp]
\includegraphics[width=1.0\linewidth]{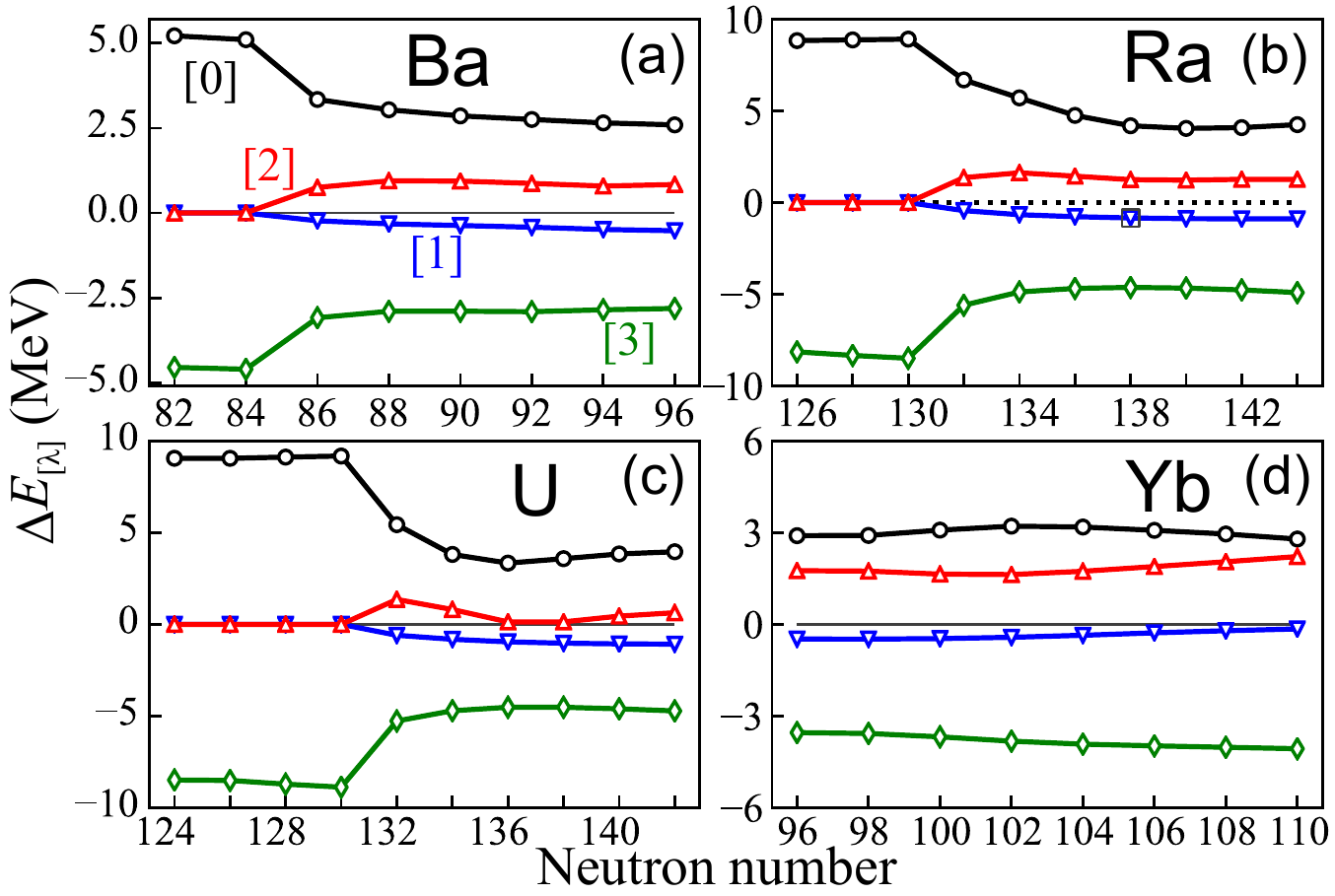}
\caption{Similar to  Fig.~\protect\ref{fig:3_isotopes} but for the  deformation
energies $\Delta E_{[\lambda]} = E_{[\lambda]}(\beta_3=0.05) - E_{[\lambda]}(\beta_3=0)$
for $\lambda=0-3$.
}
\label{fig:isotopes_multipoles}
\end{figure}
We now study $\Delta E_{[\lambda]}$, the multipole components of the
total deformation energy, for the four isotopic chains considered to see whether they could provide insights into 
the neutron-number dependence  of
octupole deformations. Figure~\ref{fig:isotopes_multipoles}  shows that
the answer is far from obvious. Indeed, we observe strong
cancellations of contributions coming from different multipole
components of the reflection-asymmetric deformation energy. For example, both the repulsive monopole
and attractive octupole components are an order of magnitude larger
than the total deformation energies shown in
Fig.~\ref{fig:3_isotopes_beta3}. Therefore,
we can  expect that in order to understand the behavior of the 
deformation energies,  higher-order multipole components $\Delta E_{[\lambda]}$ should be considered. Indeed, it has been early recognized  that higher-order deformations can strongly influence the  octupole collectivity of reflection-asymmetric nuclei
\cite{Rozmej1988,Sobiczewski1988,Egido1989,Egido1990,Cwiok1989,Cwiok1989PLB,Cwiok1991,Nazarewicz92}.

\begin{figure}[htbp]
\includegraphics[width=1.0\linewidth]{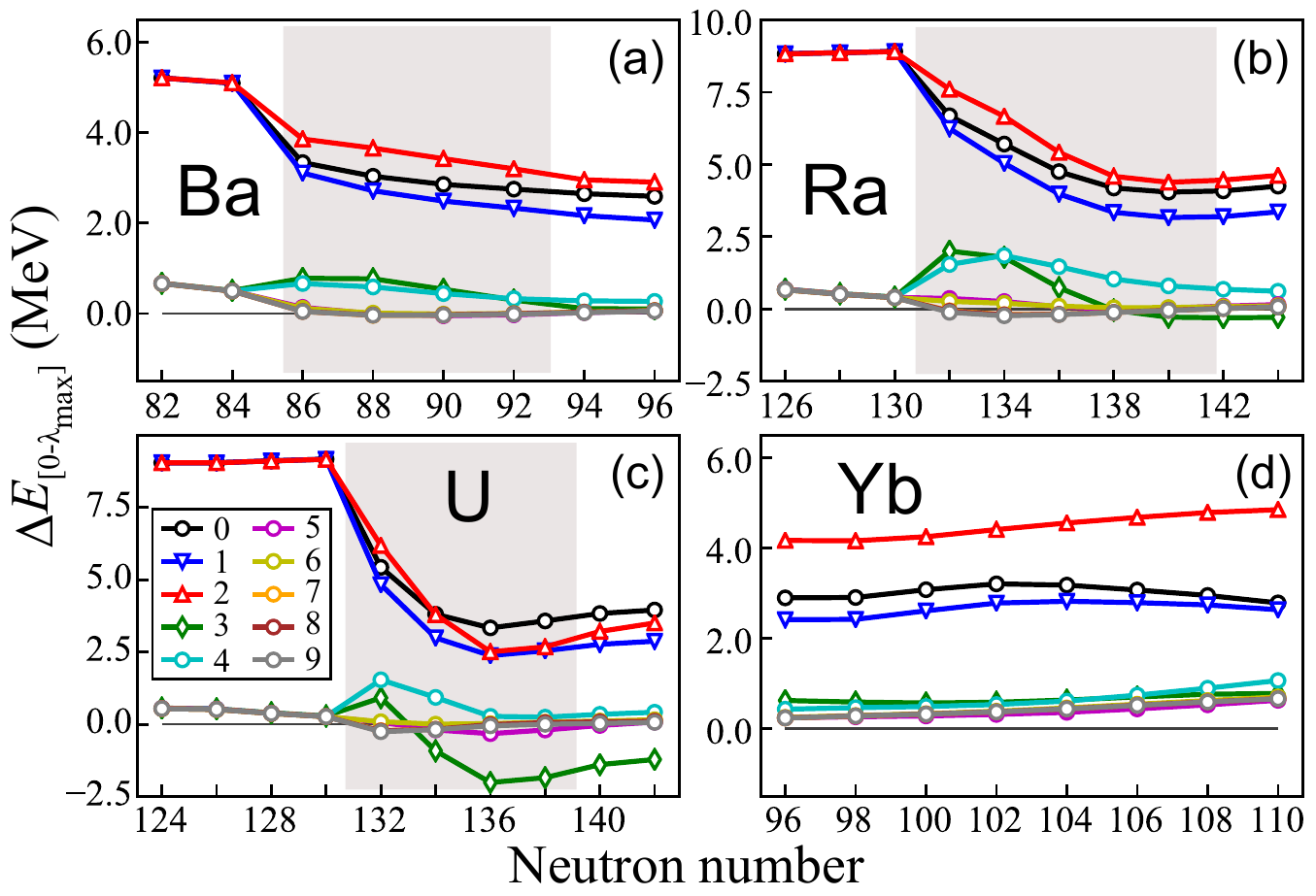}
\caption{Similar to   Fig.~\ref{fig:isotopes_multipoles} but for the 
deformation energies $\Delta E = E(\beta_3=0.05) - E(\beta_3=0)$ with multipole components summed  up from $\lambda=0$ to $\lambda_{\rm max}$. The values of $\lambda_{\rm max}$ are listed in the legend. The regions of deformed isotopes exhibiting reflection-asymmetric instability in Fig.~\ref{fig:3_isotopes_beta3} are marked by shading.
}
\label{fig:isotopes_multipoles_summed}
\end{figure}
\begin{figure}[htbp]
\includegraphics[width=1.0\linewidth]{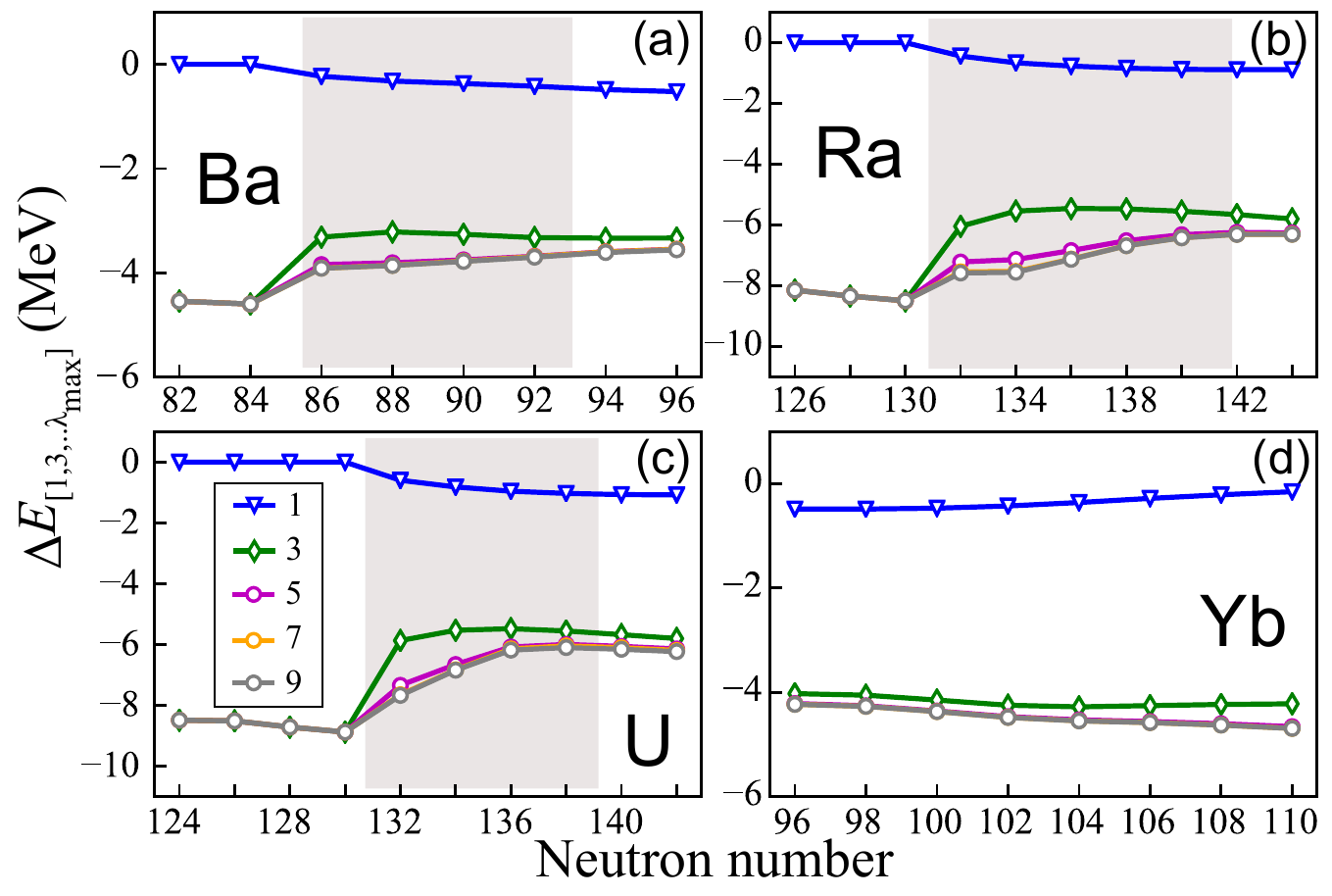}
\caption{Similar to Fig.~\ref{fig:isotopes_multipoles_summed} but for the
cumulative sum involving
 odd-$\lambda$ multipoles only.
}
\label{fig:3_isotopes_multipoles_summed_odd}
\end{figure}
\begin{figure}[htbp]
\includegraphics[width=1.0\linewidth]{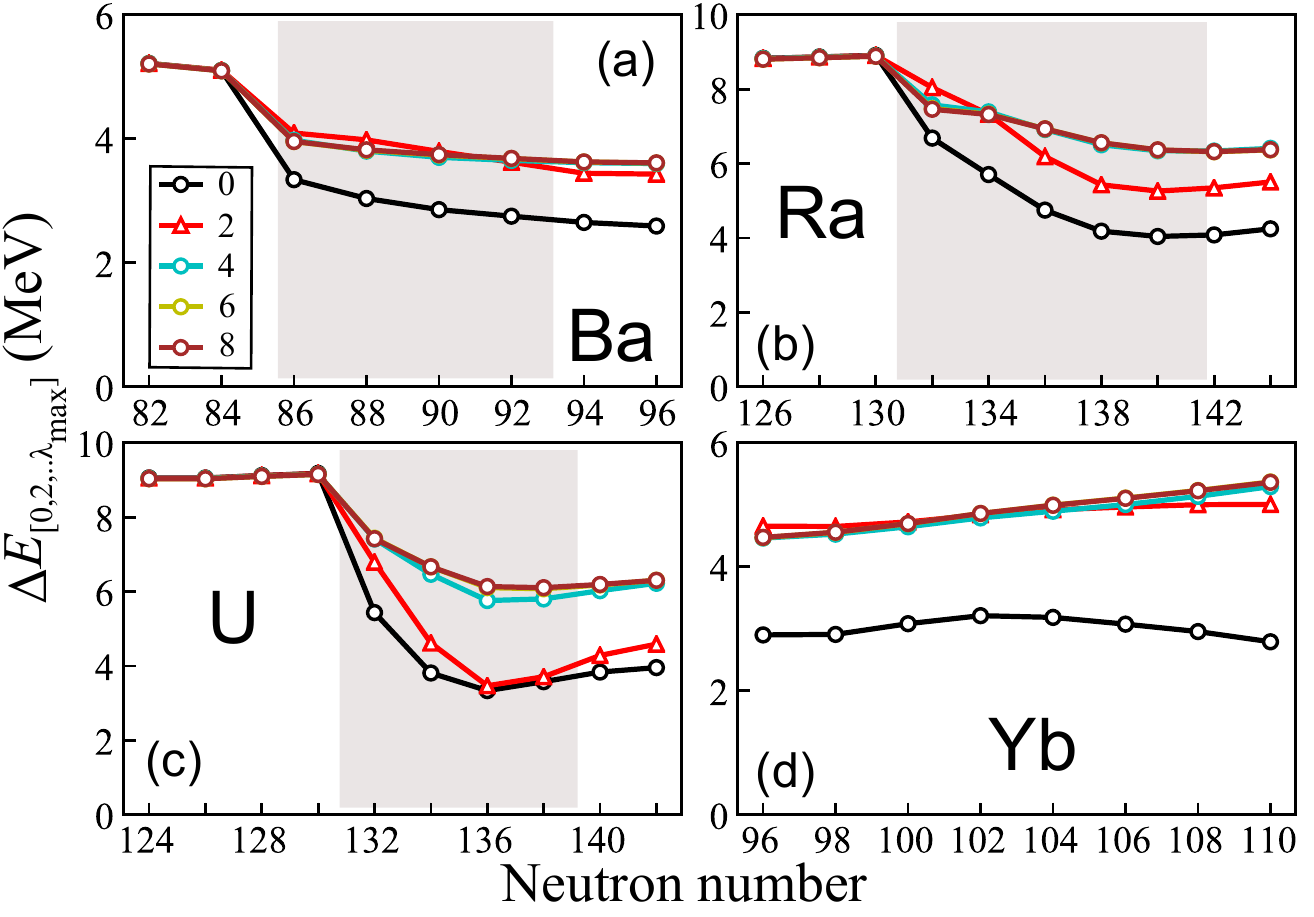}
\caption{Similar to  Fig.~\protect\ref{fig:isotopes_multipoles_summed} but for  the cumulative sum involving
 even-$\lambda$ multipoles only.
}
\label{fig:3_isotopes_multipoles_summed_even}
\end{figure}

To better see accumulation effects with increasing multipolarity
and subtle fluctuations at different orders, in
Fig.~\ref{fig:isotopes_multipoles_summed} we plot multipole
components of the octupole deformability summed up to $\lambda_{\rm max}$. Noting
dramatically different scales of Figs.~\ref{fig:3_isotopes_beta3}
and~\ref{fig:isotopes_multipoles_summed}, we see that summations up
to about $\lambda=5$ or 7 are needed for the results to converge. Although
the octupole component contributes by far most to the creation of the
reflection-asymmetric deformation   energy, its effect is counterbalanced by a very large
monopole component and, therefore, even higher multipole components
are instrumental in determining the total reflection-asymmetric deformability.
This aspect is underlined in the results shown in
Figs.~\ref{fig:3_isotopes_multipoles_summed_odd}
and~\ref{fig:3_isotopes_multipoles_summed_even}, where we separately
show analogous sums of only odd-$\lambda$ (odd parity) and even-$\lambda$ (even
parity) components, respectively. It is clear that the octupole
polarizability is a result of a subtle balance between positive
(repulsive) effect of the even-parity multipoles and negative
(attractive) effect of the odd-parity multipoles.

\begin{figure}[htb]
\includegraphics[width=1.0\linewidth]{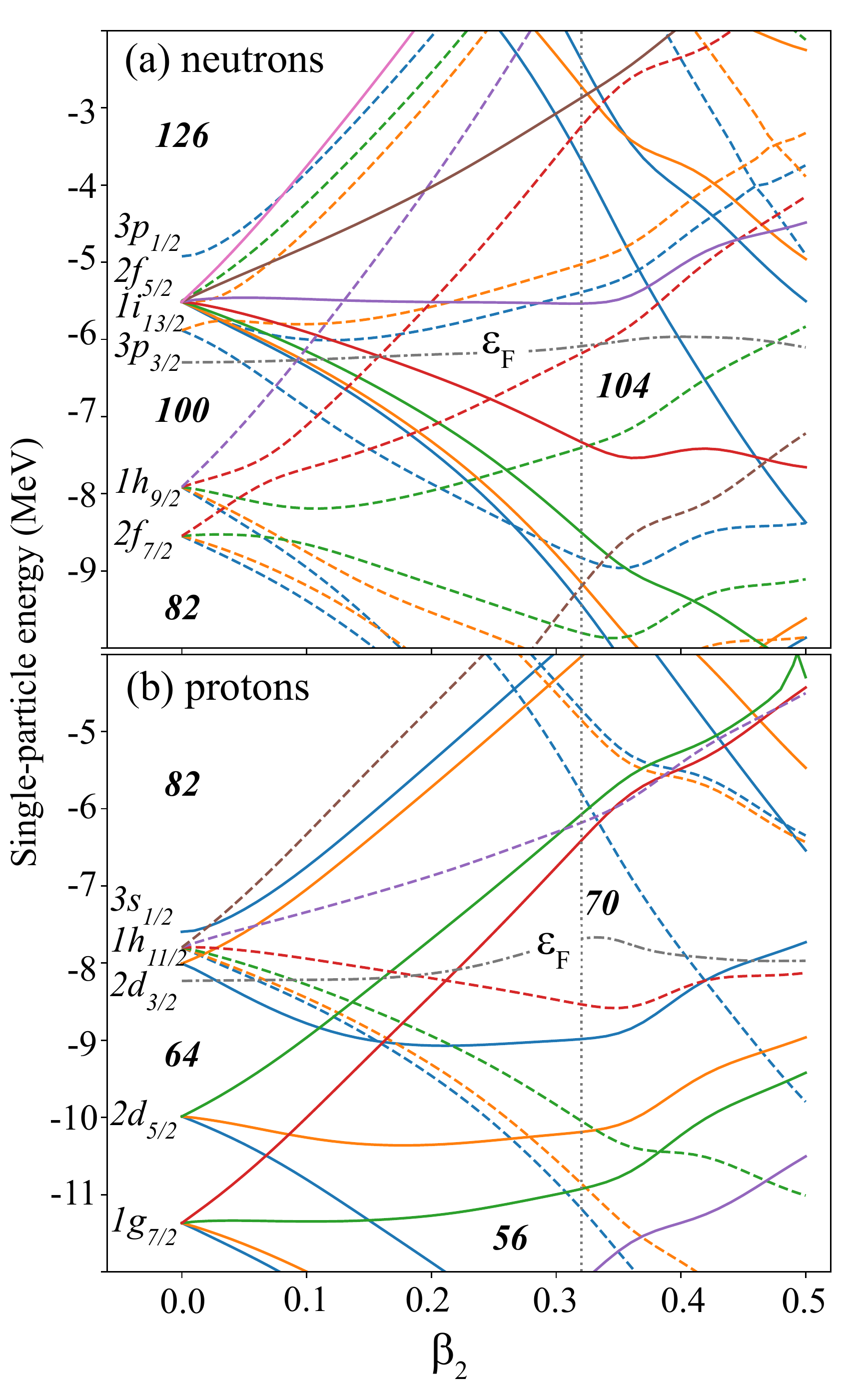}
\caption{Single-particle (canonical) neutron (top) and proton (bottom) SLy4-HFB levels as functions of $\beta_2$ ($\beta_3=0$) for $^{176}$Yb. Solid (dashed) lines indicate positive- (negative-) parity levels. Fermi levels $\varepsilon_{\rm F}$ at $N=106$ and $Z=70$ are marked by dash-dotted lines. The equilibrium deformation of $^{176}$Yb is indicated by a vertical dotted line.
}
\label{fig:Yb-levels}
\end{figure}

\subsection{Relation to shell structure}\label{sec:shells}

\begin{figure}[htb]
\includegraphics[width=1.0\linewidth]{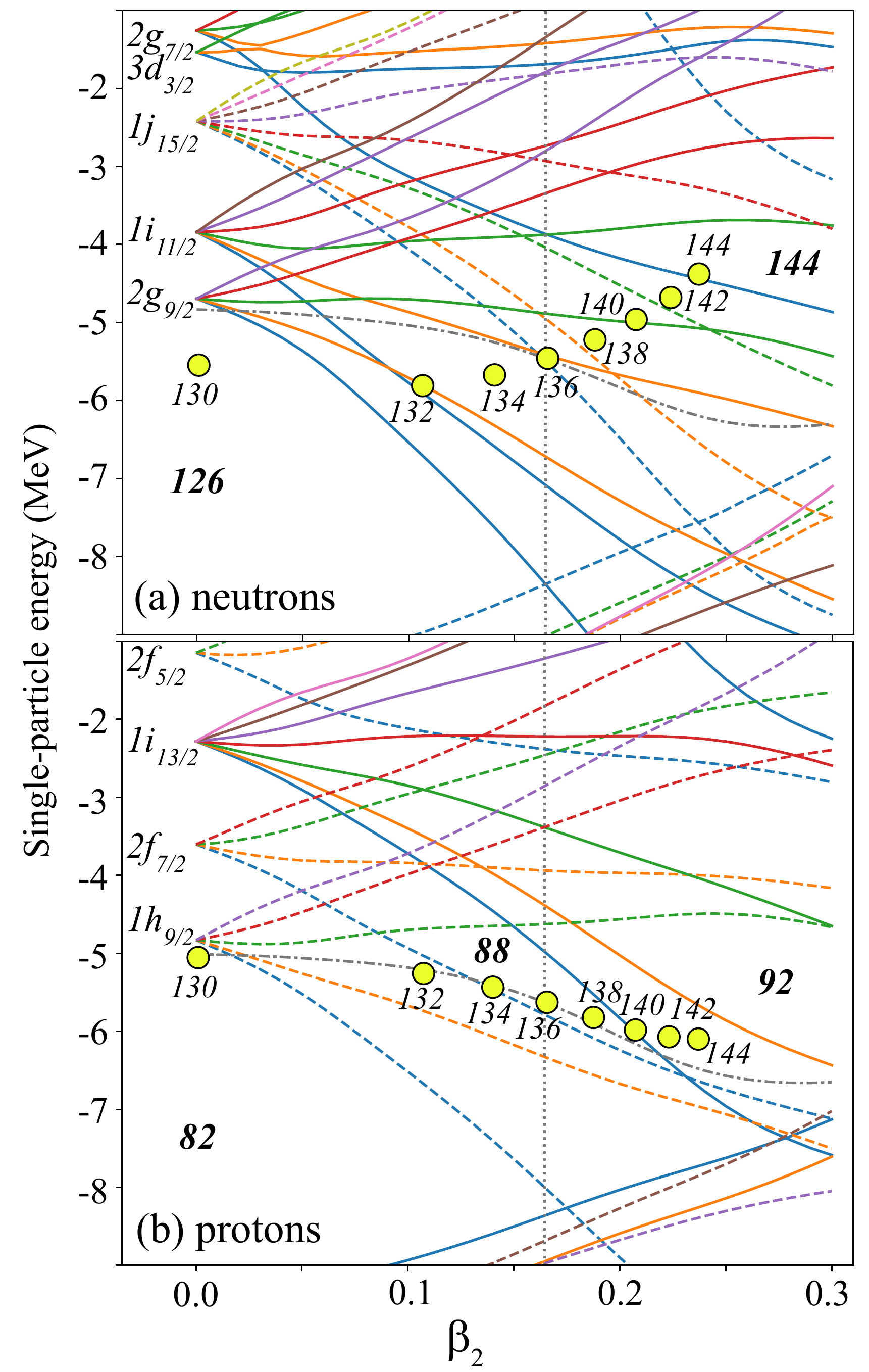}
\caption{Similar to  Fig.~\protect\ref{fig:Yb-levels} but for $^{224}$Ra.  Fermi levels for even-even Ra isotopes with $N=130-144$ are marked by circles. 
They have been shifted according to the position of the spherical 
 $2g_{9/2}$  neutron and
	  $1h_{9/2}$  proton shell.
The equilibrium deformation of $^{224}$Ra is indicated by a vertical dotted line.
}
\label{fig:Ra-levels}
\end{figure}

To gain some insights into the shell effects behind the appearance of stable reflection-asymmetric nuclear shapes, Figs.~\ref{fig:Yb-levels} and \ref{fig:Ra-levels} show, respectively,  the s.p.\ level diagrams
for $^{176}$Yb and  $^{224}$Ra as functions of $\beta_2$. While such diagrams cannot predict symmetry breaking effects {\it per se}, they  can often provide qualitative understanding.

The well-deformed nucleus $^{176}$Yb is characteristic of a stiff octupole vibrator. Indeed, its nucleon numbers ($Z=70, N=106$) lie far from the ``octupole-driving" numbers $N_{\rm oct}$. Due to the large deformed $Z=70$ gap   around $\beta_2=0.32$, there are no  s.p.\ states of opposite parity and the same
 projection $\Omega$ of the total s.p.\ angular momentum on the symmetry axis that could produce p-h excitations with appreciable $\lambda=3$ strength across the Fermi level. As for the neutron s.p.\ levels,  the low-$\Omega$ positive-parity states originating from the $1i_{13/2}$ shell lie below the Fermi level, which appreciably reduces the 
 $1i_{13/2}\leftrightarrow 2f_{7/2}$ strength. Because of the large quadrupole deformations of Yb isotopes considered, the s.p.\ orbital angular momentum $\ell$ of normal-parity orbitals is fairly fragmented within the shell \cite{Bengtsson1989}. As seen in Figs.~\ref{fig:3_isotopes_multipoles_summed_even}d
 and \ref{fig:3_isotopes_multipoles_summed_odd}d, all multipole components of $\Delta E$ for $^{176}$Yb vary very smoothly with neutron number.
 
 The Nilsson diagram shown in Fig.~\ref{fig:Ra-levels} is characteristic of transitional neutron-deficient actinides in which the  octupole instability is expected. The unique-parity shells, $1i_{13/2}$ proton shell and  $1j_{15/2}$ neutron shell, are of particle character, which results in an appearance of close-lying opposite-parity pairs of Nilsson levels
 with the same low $\Omega$-values  at  intermediate quadrupole deformations. These levels  can interact via the octupole field, with the dominant    $\pi 1i_{13/2}\leftrightarrow \pi 2f_{7/2}$ and $\nu 1j_{15/2}\leftrightarrow \nu 2g_{9/2}$ couplings.

 As seen in  Figs.~\ref{fig:3_isotopes_multipoles_summed_odd} and \ref{fig:3_isotopes_multipoles_summed_even},
in the regions of octupole instability, the monopole and quadrupole deformation energies become locally reduced while the octupole and dotriacontapole ($\lambda=5$) contributions to $\Delta E$ grow. 
According to our results, the effect of the  dotriacontapole term is essential for lowering $\Delta E$ around  $N_{\rm oct}$. This not surprising as  the main contribution to the dotriacontapole coupling comes from the $\Delta\ell=\Delta j=3$ excitations~\cite{Cwiok1989,Cwiok1991}, i.e., 
the octupole  and dotriacontapole correlations are driven by the same shell-model orbits. Interestingly, it is the attractive  $\lambda=5$ contribution to $\Delta E$ rather than the octupole term that exhibits the local enhancement in the regions of octupole instability.

 The shallow octupole minima predicted around $^{146}$Ba result from an interplay between the odd-$\lambda$ deformation energies, which gradually increase with $N$ (see 
 Fig.~\ref{fig:3_isotopes_multipoles_summed_odd}a) and the even-$\lambda$ deformation energies, which gradually decrease with $N$ (see 
 Fig.~\ref{fig:3_isotopes_multipoles_summed_even}b). Again, the dotriacontapole moment 
 is absolutely essential for forming the octupole instability.

\section{Conclusions}\label{sec:summary}

In this work, we  used the Skyrme-HFB approach to study  the multipole expansion of interaction energies in both isospin and neutron-proton schemes in order to  analyze  their role in the appearance of reflection-asymmetric g.s.\ deformations. 
The main conclusions and results of our study can be summarized
as follows:
\begin{enumerate}[label=(\roman*)]
\item
Based on the self-consistent HFB theory, reflection-asymmetric ground-state shapes of atomic nuclei are driven by the  odd-multipolarity isoscalar (or, in neutron-proton scheme, $np$)   part of the nuclear interaction energy. In a simple particle-vibration picture, this can be explained in terms of the very large isoscalar octupole polarizability  
$\chi_{3,0}(\Delta{\cal N}=1)=3$.
\item
The most favorable conditions for reflection-asymmetric shapes are in the regions of transitional nuclei with neutron and proton numbers just above magic numbers. For such systems, the unique-parity shell has a particle character, which creates favorable conditions for the   enhanced $\Delta\ell=\Delta j=3$ octupole  and dotriacontapole couplings.

\item
The presence of high-multipolarity interaction components, especially $\lambda=5$ are crucial for the emergence of stable  reflection-asymmetric shapes.
Microscopically,  dotriacontapole couplings primarily come from the same $\Delta\ell=\Delta j=3$ p-h excitations that are responsible for octupole instability. According to our calculations,
the attractive  $\lambda=5$ contribution to the octupole  stiffness  is locally enhanced in the regions of reflection-asymmetric g.s.\ shapes. 

\end{enumerate}

In summary, stable pear-like g.s.\ shapes of atomic nuclei result from a dramatic cancellation between even- and odd-multipolarity components of the nuclear binding energy.
Small variations in these components, associated, e.g.,  with the s.p.\ shell structure, can thus be instrumental for tilting the final energy balance towards or away from the octupole instability. One has to bear in mind, however, 
that the shell effect responsible for the spontaneous breaking of intrinsic parity is weak, as it is associated with the appearance of isolated  $\Delta\ell=\Delta j=3$ pairs of levels  (parity doublets) in the reflection-symmetric s.p.\ spectrum. In this respect, the breaking of the intrinsic spherical symmetry in atomic nuclei (presence of  ellipsoidal deformations) is very common as every spherical s.p.\ shell (except for those with $j=1/2$) carries an intrinsic quadrupole moment that can contribute to the vibronic coupling.

\begin{acknowledgements}
Computational resources were provided by the Institute for Cyber-Enabled Research at Michigan State University. 
This material is based upon work supported by the U.S.\ Department of Energy, Office of Science, Office of Nuclear Physics under award numbers DE-SC0013365 and DE-SC0018083 (NUCLEI SciDAC-4 collaboration); by the STFC Grant Nos.~ST/M006433/1
and~ST/P003885/1; and by the Polish National Science Centre under Contract No.~2018/31/B/ST2/02220.
\end{acknowledgements}

%

\end{document}